\documentclass{ws-ijmpa}

\begin{document}

\markboth{Doncheski, Godfrey, and Zhu}
{Higgs Boson Production in $\gamma\gamma$ Collisions via Resolved 
Photon Contributions}

%
\catchline{}{}{}{}{}
%

\title{HIGGS BOSON PRODUCTION IN $\gamma\gamma$ COLLISIONS VIA 
RESOLVED PHOTON 
CONTRIBUTIONS\footnote{Supported in part by the Natural Sciences and 
Engineering Research Council of Canada}
}

\author{\footnotesize STEPHEN GODFREY 
and SHOU-HUA ZHU\footnote{Address after December 1 2004: 
Institute of Theoretical Physics,
School of Physics, Peking University, Beijing 100871, P.R. China}
}

\address{Ottawa-Carleton Institute for Physics, Department of Physics,\\
Carleton University, Ottawa, Canada K1S 5B6 }

\author{M.A. DONCHESKI}

\address{Department of Physics, Pennsylvania State University\\
Mont Alto, PA 17237 USA Group, Laboratory, Address}

\maketitle


\begin{abstract}
We studied single Higgs boson production in $\gamma\gamma$ collisions 
proceeding via the hadronic content of the photon.  
For SM Higgs masses of current theoretical interest, the resolved photon 
contributions are non-negligible in precision cross section 
measurements.
We found that production of the 
heavier Higgs bosons $H^0$ and $A^0$ of the MSSM can probe regions of the 
SUSY parameter space that will complement other measurements.  
Finally, we showed that associated $t H^\pm$ 
production in $\gamma\gamma$ collisions
can be used to make an accurate determination of 
$\tan\beta$ for low and high $\tan\beta$ by precision measurements of 
the $\gamma\gamma \to H^\pm t +X $ cross section.  

\end{abstract}

\section{Introduction}	
The photon-photon ``Compton-collider'', which utilizes laser light
backscattered  off of highly energetic and possibly polarized electron beams, 
has been advocated as a valuable part of the LC physics program \cite{telnov}. 
We explored various aspects of Higgs boson production 
via the hadronic content of the photon\cite{Doncheski:2001uh,Doncheski:2003te}
which we briefly describe.  Details and complete references are given 
in Ref. 2 and 3. 

In the resolved photon approach the quark and gluon content of the 
photon are treated as partons described by partonic distributions, 
$f_{q/\gamma}(x,Q^2)$, in direct analogy to partons inside hadrons \cite{fph}. 
The parton subprocess cross sections
are convoluted with the parton distributions to 
obtain the final cross sections.  These have to be further
convoluted with the photon energy distributions obtained from either 
the backscattered laser or the 
Weizs\"{a}cker Williams distributions to obtain cross sections that can 
be compared to experiment.  

\section{Single Neutral Higgs Production}
One of the strongest motivations for the Compton collider 
is to measure Higgs boson properties.  
Measurement of the $\gamma\gamma \to H$ cross section 
is especially interesting as it 
proceeds via loop contributions and is therefore sensitive to new 
particles that cannot be produced directly.  It is 
important that all SM contributions to this cross section be 
carefully considered.  
The resolved photon contributions to Higgs production are shown in Fig. 1
for the backscattered laser case with $\sqrt{s_{ee}}=500$~GeV.
We also show the $\sigma(\gamma\gamma\to H)$ production which proceeds via 
loops and the contribution from 
gluon fusion, $\hat{\sigma}(gg\to H)$, which arises from the gluon content of the 
photon.  A final process  is $\gamma\gamma\to H W^+W^-$ whose
cross section is comparable to the resolved photon processes.
Although the loop process dominates over the resolved photon processes
for the full range of Higgs masses,
the 
latter processes contribute at the percent level 
for $M_H\sim 150$~GeV and $\sqrt{s_{ee}}=500$~GeV increasing
to several percent for $\sqrt{s_{ee}}=1.5$~TeV.  
Thus, 
Higgs production via the hadronic 
content of the photon may not be
negligible for precision measurement of $\sigma(\gamma\gamma \to H)$
suggesting that these contributions deserve further study.

\begin{figure}
\begin{center}
\begin{minipage}[h]{5.0cm}
\caption{Production cross sections for SM Higgs boson 
including the backscattered laser spectrum.  The solid 
line is for $\gamma\gamma\to h$, the short dashed line for 
$\hat{\sigma}(b\bar{b}\to h) +\hat{\sigma}(c\bar{c}\to h)$,
the dot-dot-dashed line for $\hat{\sigma}(c\bar{c}\to h)$, 
the long-dashed line for $\hat{\sigma}(b\bar{b}\to h)$,
the dotted line for $\hat{\sigma}(gg\to h)$ 
and the dot-dashed line for $\hat{\sigma}(WW\to H)$.} \hfill
\end{minipage} \
\begin{minipage}{7.4cm}
\centerline{\epsfig{file=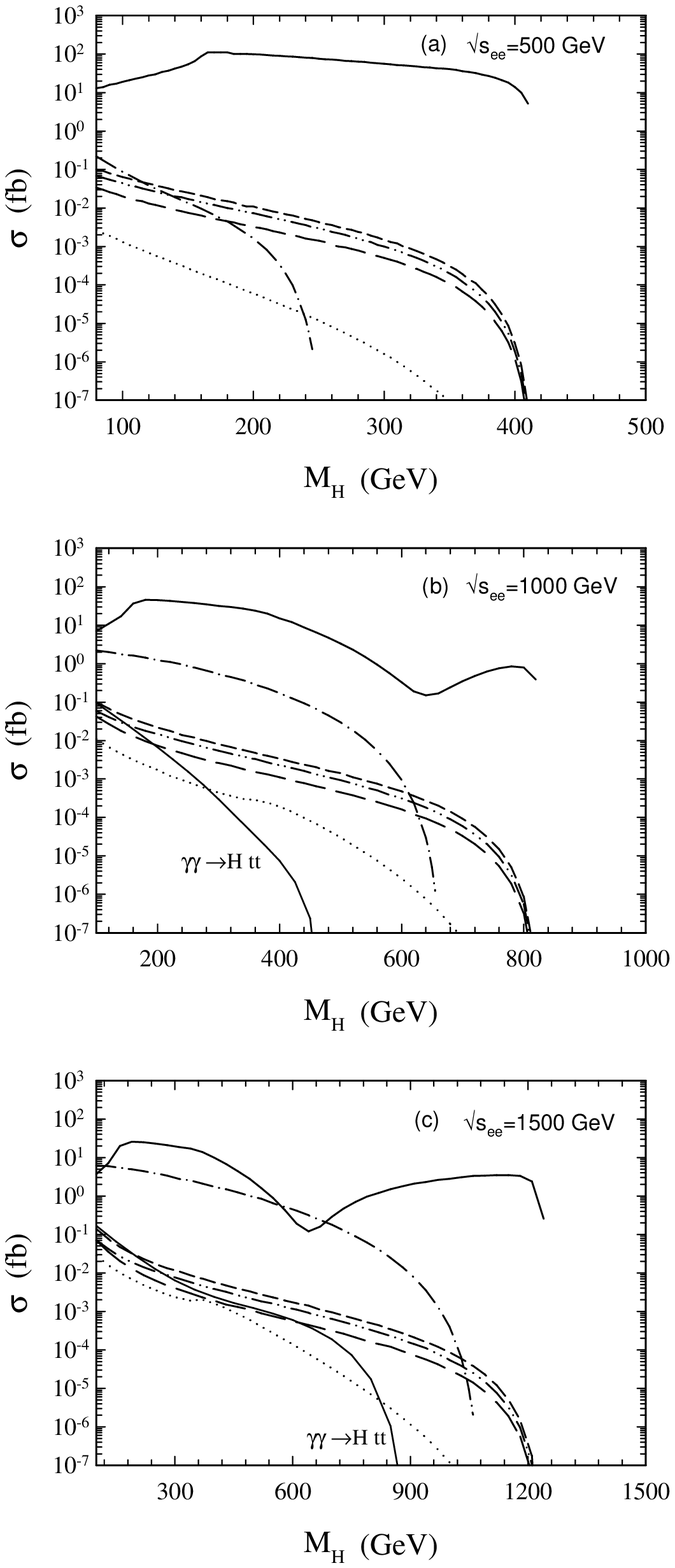,width=7.4cm,clip=}}
\end{minipage} 
\end{center}
\end{figure}


In the MSSM there exist a total of three neutral Higgs 
bosons which can be produced in 
$\gamma\gamma$ collisions; $\gamma \gamma \to h^0, \; H^0, \; A^0$.
However, the resolved 
photon process has different dependence on $\tan\beta$ than that of 
the loop processes.
If $H$ and $A$ were produced in sufficient quantity 
the cross section could be used to constrain $\tan\beta$.
This can be seen most clearly in Fig. 2 
which shows the regions of the 
$\tan\beta - M_{A}$ plot which can be explored via $A$ production
for $\sqrt{s}_{ee}=500$~GeV.  The 
regions covered would complement measurements made in other processes.

\begin{figure}
\begin{center}
\begin{minipage}[h]{6.0cm}
\caption{
Regions of sensitivity in $\tan\beta -M_A$ parameter space to $A$ 
production via resolved photons with backscattered laser photons
for $\sqrt{s_{ee}}=500$~GeV. 
The region to the left is accessable and the region to the right is 
unaccessable. 
The dotted line gives  $\sigma = 0.1$~fb contour so that at least 
100~events would be produced in the region to the left 
for 1~ab$^{-1}$ integrated luminosity.
Similary the dashed line 
gives the $\sigma = 0.02$~fb contour  designating the 
the boundary for producing at least $20$~events. 
}
\end{minipage} \
\begin{minipage}{6.0cm}
\centerline{\epsfig{file=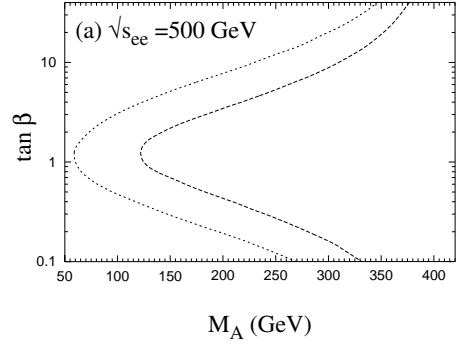,height=6.0cm,angle=-90}}
\end{minipage} 
\end{center}
\end{figure}

\section{Measurement of $\tan\beta$ in associated $t H^\pm$ Production 
in $\gamma\gamma$ Collisions}
The ratio of neutral Higgs field vacuum expectation values, 
$\tan\beta$, is a key parameter needed to be determined in 
type-II Two-Higgs Doublet Models and the MSSM. 
$\tan\beta$ can be measured in associated $tH^\pm$ 
production in $\gamma\gamma$ collisions.  The subprocess 
$b\gamma \to H^- t$ utilizes the $b$-quark content of the photon. 
$\tan\beta$ enters through the $tbH^\pm$ vertex.
In Fig. 3 we show the 
cross section as a function of $\tan\beta$ with the measurement 
precision superimposed.  
We found that  $\gamma\gamma\to tH^\pm +X$ can 
be used to make a good determination of $\tan\beta$ for most of the parameter 
space with the exception of the region around $\tan\beta\simeq 7$
where the cross section is at a point of inflection.  This measurement 
provides an additional constraint on $\tan\beta$ which 
complements other processes.  Subsequently Choi {\it et al} \cite{Choi:2004ne}
studied the
related process of $\tau\tau \to H$ where the $\tau$'s come from the 
fermionic content of the photon.

\begin{figure}
\begin{center}
\begin{minipage}[h]{5.0cm}
\caption{
$\sigma(\gamma\gamma\to t H^- +X)$ {\it vs.} $\tan\beta$ for the 
backscattered laser case and the sensitivities to $\tan\beta$ based 
only on statistical errors (solid lines) for $\sqrt{s}_{ee}=1$~TeV and 
$M_H = 200$~GeV.  For the cross sections, the solid 
line represents the expected cross section at the nominal value of 
$\tan\beta$, while the dashed (dotted) line represents the expected cross 
section at $\tan\beta - \Delta\tan\beta$ ($\tan\beta + \Delta\tan\beta$).} 
\end{minipage} \ 
\begin{minipage}[h]{7.0cm}
\centerline{\epsfig{file=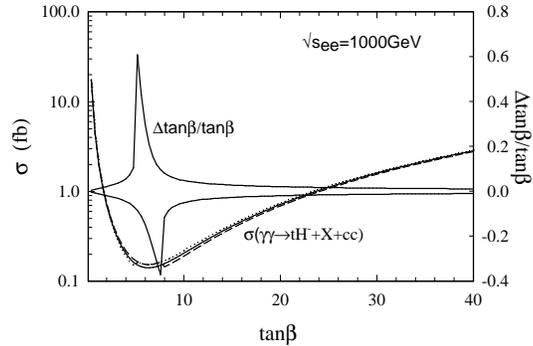,width=7.0cm,clip=}}
\end{minipage} 
\end{center}
\end{figure}

\end{document}